\documentclass[useAMS,usenatbib]{mn2e}
\usepackage[normalem]{ulem}

\usepackage{graphicx}
\usepackage{times}

\usepackage{amsmath}
\usepackage{amssymb}
\usepackage{tabularx}
\usepackage{txfonts}
\usepackage{wasysym}
\usepackage{ragged2e}
\usepackage{natbib}
\usepackage[english]{babel}
\usepackage{float}

\usepackage{color}

\usepackage{aas}

\title[Variable stars in the open clusters M\,35 \& NGC\,2158]{
  Variable stars in two open clusters within the
  \textit{Kepler\,2-Campaign-0} field: M\,35 and NGC\,2158\thanks{ Based
    on observations collected with the Schmidt\,67/92 Telescope at the
    Osservatorio Astronomico di Asiago, which is part of the
    Osservatorio Astronomico di Padova, Istituto Nazionale di
    Astrofisica.  }\thanks{Light curves of variable stars available at http://groups.dfa.unipd.it/ESPG/aphn.html\,. }}

\author[Nardiello et al.]{D.\ Nardiello$^{1,2,3}$\thanks{Ph.D.\ student when this work started.}\thanks{E-mail: domenico.nardiello@studenti.unipd.it\,.}, 
L.\ R.\ Bedin$^{1}$, 
V.\ Nascimbeni$^{1,2}\ddagger$, 
M.\ Libralato$^{1,2,4}\ddagger$\thanks{Visiting Ph.D. student at STScI under the 2013 DDRF program.},
A.\ Cunial$^{1,2}\ddagger$,
\newauthor
G.\ Piotto$^{1,2}$,
A.\ Bellini$^{4}$, 
L.\ Borsato$^{1,2}\ddagger$,
K.\ Brogaard$^{5}$,
V.\ Granata$^{1,2}$, 
L.\ Malavolta$^{1,2}\ddagger$,
\newauthor 
A.\ F.\ Marino$^{3}$, 
A.\ P.\ Milone$^{3}$, 
P.\ Ochner$^{1}$, 
S.\ Ortolani$^{1,2}$, 
L.\ Tomasella$^{1}$, 
M.\ Clemens$^{1}$, 
\newauthor
and M. Salaris$^{6}$\\
$^{1}$Istituto Nazionale di Astrofisica - Osservatorio Astronomico di Padova, Vicolo dell'Osservatorio 5, Padova, IT-35122 \\
$^{2}$Dipartimento di Fisica e Astronomia ``Galileo Galilei'', Universit\`a di Padova, Vicolo dell'Osservatorio 3, Padova IT-35122 \\
$^{3}$Research School of Astronomy and Astrophysics, The Australian National University, Cotter Road, Weston, ACT, 2611, Australia \\
$^{4}$Space Telescope Science Institute, 3700 San Martin Drive, Baltimore, MD 21218, USA \\
$^{5}$Stellar Astrophysics Centre, Department of Physics and Astronomy, Aarhus University, Ny Munkegade, 8000 Aarhus C, Denmark\\
$^{6}$ Astrophysics Research Institute, Liverpool John Moores
University, 146 Brownlow Hill, Liverpool L3 5RF, UK
}
\begin{document}

\date{Accepted 2014 December 17. Received 2014 December 16; in original form 2014 December 9; compiled \today}

\pagerange{\pageref{firstpage}--\pageref{lastpage}} \pubyear{2014}

\maketitle

\label{firstpage}

\begin{abstract}
We present a multi-year survey aimed at collecting \textit{(1)\/}
high-precision ($\sim$5 milli-mag), \textit{(2)\/} fast-cadence
($\sim$3 min), and \textit{(3)\/} relatively long duration ($\sim$10
days) multi-band photometric series. The goal of the survey is to
discover and characterize efficiently variable objects and
exoplanetary transits in four fields containing five nearby open
clusters spanning a broad range of ages.  More in detail, our project
will \textit{(1)} constitute a preparatory survey for HARPS-N@TNG,
which will be used for spectroscopic follow-up of any target of
interest that this survey discovers or characterizes, \textit{(2)}
measure rotational periods and estimate the activity level of targets
we are already monitoring with HARPS and HARPS-N for exoplanet transit
search, and \textit{(3)} long term characterization of selected
targets of interest in open clusters within the planned \emph{K2}
fields.  In this first paper we give an overview of the project, and
report on the variability of objects within the first of our selected
fields, which contains two open clusters: M\,35 and NGC\,2158.  We
detect 519 variable objects, 273 of which are new discoveries, while
the periods of most of the previously known variables are considerably
improved.
\end{abstract}


\section{Introduction}
\label{intro}
We present the first report of the multi-year, multi-wavelength
photometric survey programme ``The Asiago Pathfinder for HARPS-N'',
aimed at characterizing variable stars and transiting-exoplanet
candidates in five open clusters (OCs).  This survey will also select
promising targets for accurate HARPS-N@TNG follow-up spectroscopic
observations, including exoplanet search programmes.

High precision light curves are required for point-like source 
periodic variability searches, in particular for planetary transits. 
Although the chances of catching a transiting planet among the members of a 
typical sparse OC are rather small \citep{2011ApJ...729...63V}, one 
has to take into account that more than 50\,000 stars are detected by 
our instrument in a typical low Galactic latitude field.  This 
number has to be multiplied by the number of fields monitored during 
the survey.

Many of the clusters to be studied over the next four years are within
the field to be observed by the {\it Kepler-2 Mission} ({\it K2\,})
campaigns\footnote{http://keplerscience.arc.nasa.gov/K2/}
\citep{2014PASP..126..398H}, following the huge success of the
original {\it Kepler} mission \citep{2010ApJ...713L.131K}.  Therefore
the timing of our investigation is particularly important, as our
observations will be useful extending the baseline of the {\it K2}
survey ({\it K2} will observe a given field for only about two
months).  It will help to identify long-period variables (magnetic
cycles, Mira oscillations, etc.) and refine the ephemeris of every
transit-like signal which is detectable from the ground.

In this study, we focused on the search for photometric variable
sources in one of our selected fields, a patch of sky covering $\sim
0.6$ square degrees and containing both the young, sparse and
relatively nearby OC Messier\, 35 (M\,35 = NGC\,2168; age 180~Myr;
apparent distance modulus $(m-M)_V\simeq 10.4$;
\citealt{2003AJ....126.1402K}) and the older, more distant, and more
concentrated NGC\,2158 (1.9~Gyr; $(m-M)_V\simeq 14.3$;
\citealt{2010ApJ...708L..32B}).

There are several papers devoted to the search and study of variable
stars in these two OCs. For example, \citet{2009ApJ...695..679M}
presented a detailed study of the period-colour relation of the
rotating stars of M\,35, and
\citet{2004AJ....128..312M,2006AJ....131.1090M} studied the
variability in a field centred on NGC\,2158. In particular they found
a candidate transiting-exoplanet (TR1), that we are not able to
confirm in this work.

The aim of this investigation is to discover new variable stars in the
{\it K2-Campaign-0} field containing the two target OCs, and refine
the periods of the already catalogued variables. The paper is
structured as follows: in Sect.~\ref{data} we report our data-set; in
Sect.~\ref{reduction} and \ref{detrending} the software for the
extraction and detrending of light curves is presented; in
Sect.~\ref{finding} we show the tools used for finding variables
stars; Sect.~\ref{variables} is dedicated to the characterization of
the variables found in the M\,35 and NGC\,2158 field; in
Sect.~\ref{emat} we describe the available electronic
material. Finally, Sect.~\ref{summary} summarizes our work.

%%%%%%%%%%%%%%%%%%%%%%%%
%%%%%%%%%%%%%%%%%%%%%%%%
%%%%%%%%%%%%%%%%%%%%%%%%

\section{The Database}
\label{data}

All data used in this work come from the Asiago 67/92~cm Schmidt 
Telescope, located at 1370~m on Mt.\ Ekar 
(longitude 11$^\circ\!$.5710~E, latitude 45$^\circ\!$.8430~N); this 
facility belongs to the Astronomical Observatory of Padova (OAPD - 
Osservatorio Astronomico di Padova), which is part of the Istituto 
Nazionale di Astrofisica (INAF).  The instrument mounted at the 
Schmidt focus is a SBIG STL-11000M camera, equipped with a 
4050$\times$2672 pixel Kodak KAI-11000M detector, having a pixel size 
of $9\mu\textrm{m} \times 9\mu\textrm{m}$, a pixel scale of 862 mas 
pixel$^{-1}$ (resulting in a 58$^\prime\times$38$^\prime$ field of 
view), electronic gain of 0.92~e$^{-}$/ADU, and a readout noise (RON) of 
12 e$^{-}$/s.  The detector is cooled with a thermoelectric 
Peltier stage, coadiuvated by a radiator with glicole that keeps the 
operating temperature between $-30$ and $-20$ $^\circ$C. 

Under the long-term observing programme ``The Asiago Pathfinder for 
HARPS-N'' (PI:\ Bedin) four fields were awarded with 80 nights per 
year, during 3 observing campaigns.  Taking into account the weather 
losses, the number of nights statistically guarantees a minimum of ten  
nights per year in 3 observational campaigns. 

\begin{figure*}
\includegraphics[width=\hsize]{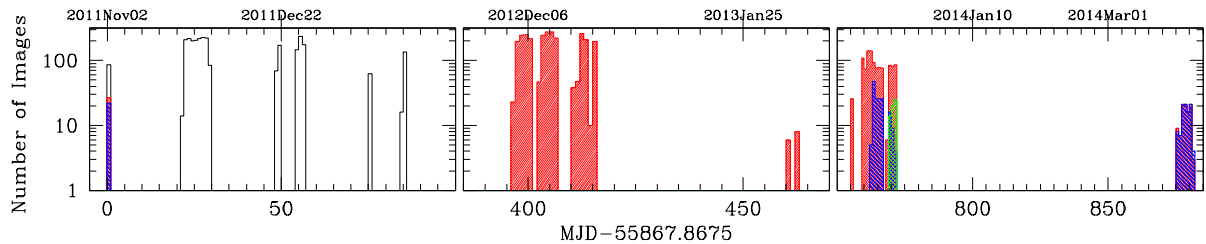}
\caption{
  Histogram of the number of images per night collected during the 
  first three campaigns of our programme.  The white histogram refers to 
  observations in white light, the red, blue and green histograms  
  refer to observations in the $R$, $B$ and $V$ filters, respectively.}
\label{obs}
\end{figure*}

The first observing season was a pilot program.  We aimed
to have the highest number of stars collected with as many
photons as possible. Therefore data were collected in white light
(hereafter indicated with filter $N$, for ``None''). The unfiltered
CCD throughput roughly peaks at $\sim$500 nm.  The exposures were
long 120\,s, long enough to maximize the duty cycle, the
cadence, and the dynamic range vs.\ sky brightness; in this way we
were able to measure faint stars (down to $V\sim20$) and to
monitor more stars in this relatively low Galactic-plane field.
After analyzing season-one data, we realized that the number of stars
was adequate, and that we would be able to follow the same stars in a
more thoughtful pass-band, filter $R$, at the cost of slightly increased
exposure times (180\,s). This would allow better comparison with
stellar models and at the same time suppress noise due to larger
chromatic and sky-brightness effects.
It was also clear from the pilot data-set that adding short exposures
of 15\,s would have extended the photometric monitoring to a
non-negligible sample of saturated stars.  There is an additional
benefit in alternating short and long exposures besides the increased
dynamic range. The use of short exposures means that unsaturated data
will still result from the bulk of the stars in the case of extremely
good seeing.  Short exposures could not be shorter than 15\,s because
of the reliability on the shutter time.
Therefore, when season two and three were accepted we decided to
alternate between these short and long exposures in filter $R$ only.

Preliminary results from first two seasons shown that we were able to
find many new variables in the field (and 21 variables were found in
the photometric series from short exposures, Sect.~\ref{finding}).

Once the variable stars are found, the natural follow up is to collect
information about their spectral energy distribution (SED).
Therefore, in the third season we collected long exposures alternating
both $B$ and $R$ and occasionally we took exposures in $V$.  Colour
information is particularly important for the characterization of the
components of binary systems, not only off-eclipse, but possibly also
during the eclipses.  
Given the successful outcome of this project, additional observing
time for follow-up observations in $I$-band has been allocated
(unfortunately $U$ is not doable from Asiago) for the M\,35 and
NGC\,2158 field.

During the first year the fine pointing of the field was not
optimal. There is an incomplete overlap between the $N$ and $B\,R\,V$
photometric series, resulting in a $\sim$10\% sky area not imaged
through all the four filters.

Observations included standard calibration data (bias, dark, sky and
dome flats) at the beginning and end of each night. In Table~\ref{tab:log} we give
a log of the observations. Figure~\ref{obs} shows the histograms of the number
of images gathered each night for a given filter during the three seasons.

For the other clusters of this survey we will adopt  the
following strategy: long + short $R$ exposure in the first season, and
multi-filter characterization in the following seasons. With our
instrumentation this turns out to be a rather effective strategy,
doubling the number of known variables in a relatively well-studied
field such as that of M\,35 and NGC\,2158.\footnote{
``A posteriori'', knowing the characteristics of the M35 and NGC2158
field, it would have been better to have the variable finding
campaign in the 3 seasons in short and long$+$short $R$, and
colour information from $B$, and occasionally $V$ and $I$.
}
%

%%%%%%%%%%%%%%%%%%%%%%%%
%%%%%%%%%%%%%%%%%%%%%%%%
%%%%%%%%%%%%%%%%%%%%%%%%
\begin{table}
\caption{Log of observations.}
\medskip
\label{tab:log}
\begin{tabular}{l c c c c}
\hline
{\bf Filter} & {\bf \# Images} & {\bf Exp. Time}    & {\bf FWHM}  & {\bf median FWHM}  \\
             &           &  (s)               &  (arcsec)   &  (arcsec)          \\
\hline 
\hline
$B$          &    21     &    $120$          & 1.44--5.28  & 2.65               \\
             &    258    &    $180$          &             &                    \\
             &    1      &    $240$           &             &                    \\
$V$          &    60     &    $180$           & 1.24--2.05  & 1.43               \\
$R$          &    1385   &    $15$           & 1.35--6.34  & 2.75               \\
             &    27     &    $120$          &             &                    \\
             &    2552   &    $180$           &             &                    \\
$N$  &    2692   &    $120$           & 1.67--7.05  & 2.85               \\
\hline
\end{tabular}

\justify
\end{table}

\section{Data Reduction}
\label{reduction}
For all our reductions we developed custom software tools written in 
\texttt{FORTRAN~77} or \texttt{Fortran~90/95}, adapted from the same 
software described in previous papers by some of the co-authors of 
this work (we will give in the following brief descriptions and 
references).

\subsection{Pre-reduction}

The first stage of our pipeline produces master biases, master darks 
and master flats for each night. Such master frames are clipped means 
of the individual calibration images gathered on each observing night. 
We almost always used dome flats, because they proved to be more 
accurate and stable than typical sky flats.  Flats were 
bias-subtracted, while science images were dark-subtracted, the master 
dark being constructed from darks having the same exposure time as the 
photometric series. 
The dark-subtracted scientific images were then divided by the 
bias-subtracted flats of the corresponding filter.  The correction was 
performed by using a custom master-flat field for each night.

\subsection{Point Spread Functions} 
\label{psfs}

The point-spread function (PSF) of the Asiago Schmidt camera is almost 
always well sampled, with the exception of a few images collected in 
the best-seeing conditions. In these cases only the central part of 
the detector is affected by undersampled PSFs, in which the image 
quality, measured as the full width at half maximum (FHWM), is below 
1.8 arcsec. 

To compute the PSF models, we developed the software 
\texttt{img2psf\_Sch} in which our PSF models are derived in a 
completely empirical fashion. This code follows a similar 
software developed for WFI@ESO/MPG 2.2m data, and described in depth in 
\citet{2006A&A...454.1029A}. The Asiago Schmidt's PSFs are modeled 
through an array of 201$\times$201 grid points, which super-sample the 
PSF pixels by a factor of 4 with respect to the image pixels.  A 
bicubic spline is then fitted to interpolate the value of the PSF in 
between these grid points.  Furthermore, in order to model the 
considerable amount of spatial variations of the PSFs across the field 
of view of each individual image, we divided the detector in 
9$\times$5 regions, and empirically derived the PSFs independently 
within each of these subregions.  A bilinear interpolation is then 
applied to obtain the best PSF model at any specific location on the 
detector.  This interpolation is performed for every individual star 
on each frame \citep{2006A&A...454.1029A}. 

\subsection{Astrometry}

\begin{figure*}
\includegraphics[width=\hsize]{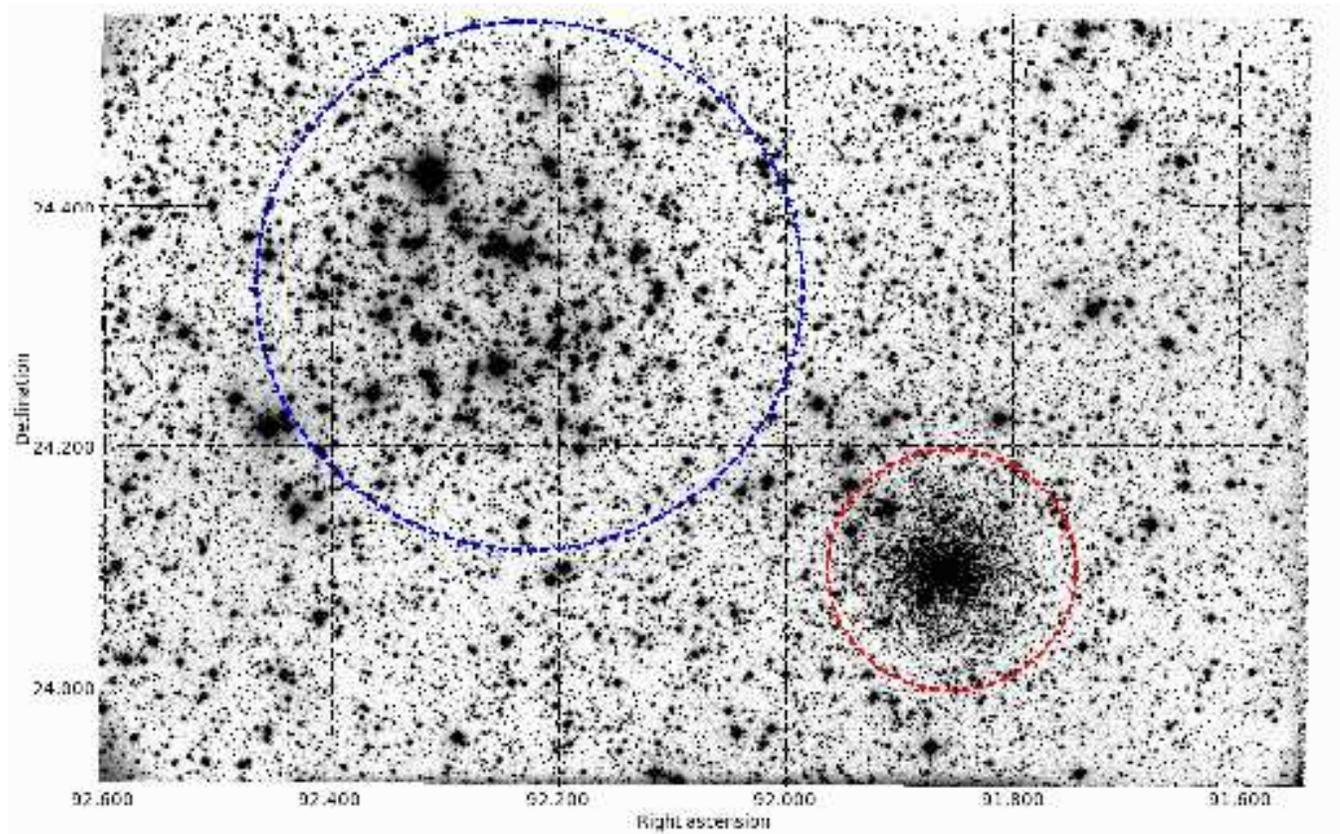}
\caption{
  Stacked $R$-filter image from our data. The blue circle marks the 
  M~35 region. The red circle marks the region of NGC~2158.  }
\label{fov}
\end{figure*}

During the first and the third observing seasons, 25 images with both
large and small dithers were collected with the purpose of solving for
the geometric distortion of the Schmidt's camera. We carried out the
same self-calibration procedure described in several of our previous
works (for example:
\citealt{2014A&A...563A..80L,2010A&A...517A..34B,2011PASP..123..622B,2006A&A...454.1029A}).
The average geometric distortion is about 1 pixel from corner to the 
centre of the detector (i.e., $\sim$1\,arcsec). 

We applied this geometric distortion solution to the raw star
positions obtained by fitting our empirical PSFs in each individual
image.  The distortion-corrected positions of each image were then
transformed into a common distortion-corrected reference frame.  We
employed as reference frame the distortion-corrected positions of the
image with the smallest airmass and the best seeing in $R$ (the ID of
this image is \texttt{SC23779}).

We considered the most-general linear transformations ( six
parameters, i.e., two shifts, rotation, a scale-factor, two skew
terms), and derived through a linear least-square fit of the pairs of
distortion-corrected coordinates of all (well-measured) stars in
common between the two frames (the considered frame, and the reference
frame). Such lists of pairs were saved in so-called transformation
files.

The consistency of positions on the reference frame tells us how well
we are able to transform the coordinate system of one image into
another image taken at a different epoch. For the best stars (below
saturation, isolated and measured with a high signal to noise ratio)
we found a consistency in position of about 20\,mas (i.e., $\sim$0.02
pixels).  Not enough to determine accurate proper motions for most
stars with the available time baseline, but accurate enough to
register time series photometry.

\subsection{Stacked Images}
\label{stack}
The transformations from the distortion-corrected coordinates of each 
image into the reference frame were used to create a stacked, high-S/N 
image of the field for each filter. The ``stack'' provides a deeper 
view of our FOV.  The stack for the filter $R$ is shown in Fig.~\ref{fov}. 
We edited the header of each of these stacked images, adding World 
Coordinate System keywords, where the absolute astrometric solution 
was computed matching the 2MASS \citep{2006AJ....131.1163S} point source 
catalogue with our sources. The absolute astrometric accuracy of 2MASS 
is estimated to be about 200\,mas. 

As part of the material provided in this paper, we make electronically 
available the astrometrized stacked images for each filter. 

\begin{figure*}
\includegraphics[width=\hsize]{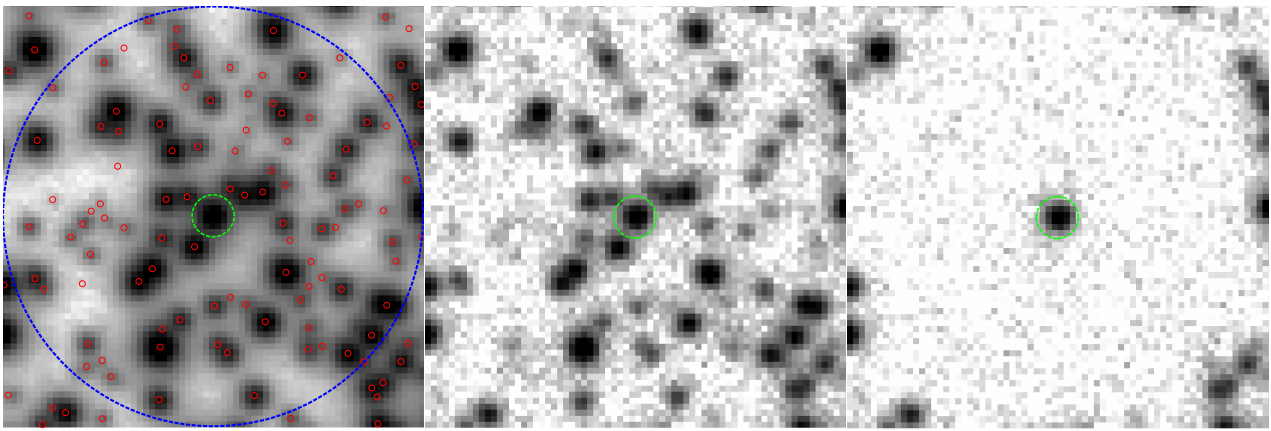}
\caption{
  An example of neighbour subtraction. The left panel shows a typical 
  subregion of the stacked image; all detected sources are marked with 
  red circles.  The middle panel shows the same for an individual image 
  (\texttt{SC23893}); the target star is located at the centre of the 
  panel (marked by a green circle). The right panel shows the 
  same region after the subtraction of neighbours within a radius of 35 
  pixels. }
\label{sub}
\end{figure*}

\subsection{The master star list}

Stacked images contain much more information than individual exposures, 
having essentially a much higher S/N.  In the case of the $R$ and $N$ 
stacks, images have integrated exposure times thousands of times 
longer than that of a single image.  The increased depth means that we 
could extract more complete and unbiased star lists.  Using the same 
software described in Sect.~3.2, we derived improved star lists from 
the $R$ and $N$ stacked images, i.e., the deepest ones.  Our finding 
algorithm takes all the local maxima above the local sky with an 
integrated flux of at least 3 digital numbers (DNs), and isolated by 
at least 3 pixels from the nearest peak.  Positions and fluxes on the 
stacked images were obtained by simultaneous PSF fitting of the target 
star and all of its neighbours, in an iterative fashion.  The program 
that performs the finding and simultaneous iterative PSF fit, 
\texttt{img2xym\_Sch}, is an adaptation of the code  
\texttt{img2xym\_WFI} presented in \citet{2006A&A...454.1029A} and 
thereby described in details. 
All the objects in the $R$ star list were then transformed to the 
astrometric and photometric system of the reference image (ID 
\texttt{SC23779}). For the $N$ star list, the photometric reference 
system of image \texttt{SC25458} was adopted, but coordinates were 
kept in the $R$ astrometric reference frame. 

The above mentioned star lists initially contained a large number of 
false detections, such as PSF artifacts, warm pixels, etc.  We purged 
the star lists from artifacts and non-stellar objects as follows. 
Most of the background galaxies were discriminated using the 
\texttt{qfit} parameter, a diagnostic related to the quality of the 
PSF fit and described in \citet{2008AJ....135.2114A}.  The artifacts 
of the PSFs are identified using the procedures described in 
\citet{2014A&A...563A..80L}. 

The final $N$+$R$ combined and purged star list contains 66\,486 
objects and constitutes the catalogue that will be used for the 
extraction of light curves throughout the following sections. 
We refer to this catalogue as Master Star List (hereafter, MSL).

\subsection{Photometry with and without neighbours} 
\label{photwithnonei}

To extract light curves, we developed a parallel code written in 
\texttt{Fortran 90/95} which uses \texttt{OpenMP}, in order to run 
simultaneously on several CPU cores (32 on our workstation).  The 
software takes as input 
\textit{(i)} the MSL catalogue as defined in the previous section, 
\textit{(ii)} the PSFs described in Sect.~3.2, and \textit{(iii)} the 
lists of pairs of coordinates described in Sect.~3.3 (transformation 
files; stars in common between the MSL and each individual image).  In 
this work we extracted the raw fluxes of all point source detected in the 
field (i.e., in the MSL) using both PSF and aperture photometry, 
keeping for the final stage of the analysis only the photometric reduction 
that minimizes the scatter in the extracted light curves (as we 
describe in the next sections).  We already described the local PSF 
technique in Sect.~3.2, while aperture photometry is computed through 
a traditional approach, by running a software pipeline originally 
developed for the TASTE project (The Asiago Search for Transit time  
variations of Exoplanets; \citealt{2011A&A...527A..85N}). The 
underlying routines are explained in details in 
\citet{2013A&A...549A..30N}. 

For each target star in the MSL, light curves are extracted in two 
parallel versions.  A first version from the original images, and a 
second one from images where the neighbours close to the target star 
were PSF-fitted and subtracted.  The neighbour subtraction is done as 
follows. 
For each target star we computed 6-parameter local transformations
between the MSL reference systems and the distortion-corrected
coordinates of stars in the individual image, using the best measured
bright, unsaturated and isolated stars located within 500 pixels of
the target star.  The magnitudes in the two reference systems are also
used to register the fluxes in the two photometric systems.  With
these local transformations, magnitude shifts, and PSFs, the stars of
the MSL located less than 35 pixels from the targets are modelled and
subtracted from each individual image. Note that such modelling
requires an inversion of the geometric distortion solution to obtain
the coordinates on the raw image reference system.
Figure~\ref{sub} compares a patch of sky around a 
typical target star: 1) on the stacked image, 2) in an individual image 
before the subtraction of the neighbours, and 3) after the 
neighbour subtraction.
After the subtraction procedure, the software performs four parallel 
reductions: aperture and PSF-fitting photometry of the target star, 
both on the original frames and on the images where the nearby stars 
are subtracted. The centroid used for aperture photometry of the 
target star is computed using the same local 6-parameters 
transformations described above.

We found that adopting a dynamical aperture, i.e., adapting the radius
$r$ of the circular photometric aperture to each image, provides the
best photometry. From theoretical computations we know that in the
case of a two-dimensional Gaussian PSF the optimum S/N ratio is
provided by an aperture radius $r\approx 0.68\times {\rm FWHM}$ (see
for example: \citealt{1999ASPC..189...50M}). About 72\% of the total
stellar flux is enclosed within this radius.
However, we empirically found that larger apertures result in light
curves of smaller scatter for objects at the bright end of our
sample. As a compromise, we adopted $r=1\times{\rm FWHM}$, which
improves aperture photometry for stars having instrumental magnitudes
$<-10.5$, but at the cost of worsening the photometry of faintest
stars.

The fitting radius of the PSFs was set to a constant value, equal to 
2.65\,pixels, as it proved to minimize the 
scatter of the light curves at all magnitudes.  

For all the four reductions we computed the local value of the sky 
background within a circular annulus of inner radius $r_{\rm 
  in}=7.5\,\sigma$ and outer radius $r_{\rm out}=r_{\rm in}+8$, where 
$\sigma = {\rm FWHM}/2.355$ is the standard deviation of the best-fit 
Gaussian PSF.

\begin{figure*}
\includegraphics[width=\hsize]{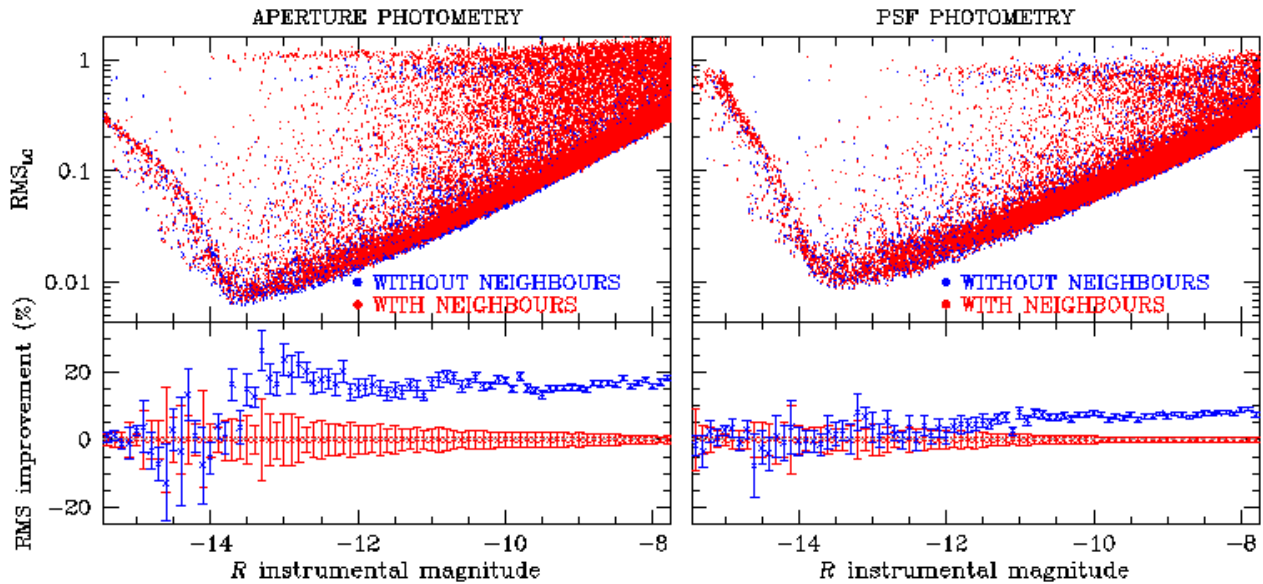}
\caption{
Photometric RMS derived from the original ``long'' $R$ images (in red) 
compared to the RMS obtained from the same images after neighbour-subtraction 
(blue), for both aperture photometry (\emph{left}) and 
 PSF photometry (\emph{right}).  The lower panels show the percentage 
improvement of the neighbour-subtracted algorithm on the RMS, averaged 
over 0.1-mag bins. The error bars are given as the 68.27th percentile 
of the residuals around the median values.
}
\label{nei_nonei}
\end{figure*}

We found that photometry on images after neighbour subtraction 
performs --on average-- better than that computed on original 
images. This is illustrated in Fig.~\ref{nei_nonei}: the top panels display  
the comparison between the photometric RMS (left panel in the case of 
aperture photometry, right panel in the case of PSF photometry) 
obtained from original images (red) and the photometric RMS after 
neighbour subtraction (blue); the lower panels show the improvement of the 
neighbour-subtracted algorithm regarding the photometric RMS, averaged over 
0.1-mag bins.  Therefore, in the following (unless otherwise quoted) 
we will make use of light curves derived from aperture and PSF 
photometry obtained from neighbour-subtracted images.

%%%%%%%%%%%%%%%%%%%%%%%%
%%%%%%%%%%%%%%%%%%%%%%%%
%%%%%%%%%%%%%%%%%%%%%%%%

\section{De-trending of Light Curves}
\label{detrending}
To remove residual systematic errors from the light curves derived in
the previous section we followed the procedure described in detail by
\citet{2014MNRAS.442.2381N}. For each target star we selected a group
of reference stars, which are used to define the local photometric
zero point of each target in all the individual images.  These
reference stars were empirically and iteratively weighted to minimize
the scatter on the final differential light curve.  A first list of
candidate reference stars is selected by computing the median
photometric scatter $\sigma$ of the raw light curves as a function of
magnitude, then binning over magnitude bins and discarding every
source with a RMS $4\sigma$ larger than the median RMS of the
corresponding magnitude bin.

A short description of the detrending algorithm follows.  Let us 
consider the $k$-th target star, the $j$-th reference star and the $i$-th 
epoch where both $k$ and $j$ stars are detected.  First we 
computed the differential light curve $m_{i,k}-m_{i,j}$, its median 
magnitude and scatter $\sigma_{jk}$, which are then used to compute the 
initial weights $w_{jk}=1/\sigma^2_{jk}$ for each reference star. The 
final weights are obtained multiplying $w_{jk}$ by two additional 
factors: $W_{jk}=(1/\sigma^2_{jk}) \cdot D_{jk} \cdot M_{jk}$. The 
factor $D_{jk}$ is an analytic function of the relative on-sky 
position ($\rho_{jk}$) between the reference star and the target star: 
it is zero within a radius $r_0=20$\,pix to avoid blending and/or 
contamination, it is equal to one between $r_0$ and $r_{\rm 
  in}=200$\,pix, and decreases exponentially with $\rho_{jk}$ from one 
to zero with a $r_{\rm out}-r_{\rm in}$ scale radius, where $r_{\rm 
  out}=300$\,pix.  The weight $M_{jk}$ is defined in a similar way, as 
a function of the magnitude difference $\phi_{jk}=\mid m_j-m_k\mid$ 
between the target and the reference star: it is equal to unity when this 
difference is less than $f_{\rm in}=1.0$\,mag, otherwise it decreases 
exponentially with $\phi_{jk}$ with $f_{\rm out}-f_{\rm in}$ scale 
radius and $f_{\rm out}=1.75$\, mag. The $D_{jk}$ and $M_{jk}$ factors 
assign a larger weight to reference stars which are closer to the target 
and of similar brightness, minimizing flux- and position-dependent 
systematics. 

\begin{figure*}
\includegraphics[width=\hsize]{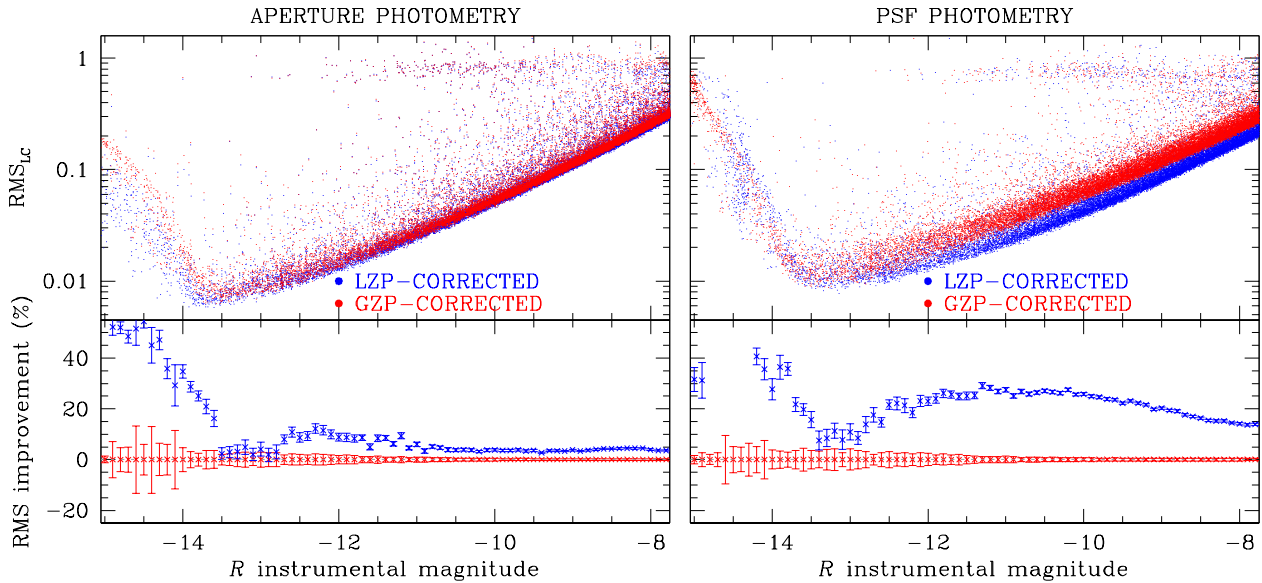}
\caption{
Photometric RMS, obtained from neighbours subtracted images, for 
GZP-corrected (red) and LZP-corrected light curves (blue).  This 
is shown for both aperture photometry (\emph{left}) and PSF photometry 
(\emph{right}). As in Fig.~\ref{nei_nonei}, the bottom panels 
show the improvement of the LZP algorithm compared to the GZP one.  }
\label{raw_detrend}
\end{figure*}

For comparison purposes, we computed two different zero-point
corrections.  A global zero point correction (GZP) $\tau_i$, defined
by the formula:

\begin{equation}
m^\prime_{i,k}=m_{i,k}-\tau_i=m_{i,k}-\langle m_{i,j}- \langle m_{i,j} \rangle_i  \rangle_j
\end{equation}

\noindent where the notation $\langle x \rangle_y$ represents the averaging of 
$x$ over the index $y$. The output of GZP is then a classical, 
unweighted differential photometry, where all the reference stars have 
unitary weight.  The local zero point correction (LZP) $\tau^\prime$ 
is computed with an expression equivalent to $\tau$, but this time 
using the weighted mean of magnitudes of our set of reference stars, 
where the weights are assigned as $W_{jk}$ and computed as above. 
Figure~\ref{raw_detrend} compares the scatter of the light curves 
corrected using the GZP (red) and LZP (blue) detrending algorithms, 
both for aperture photometry (on the left panels) and for PSF-fitting 
photometry (on the right panels). The improvement obtained with the 
LZP correction is evident (10--20\%), especially on the bright, 
non-saturated side of the sample.  For this reason, only the LZP light 
curves are analyzed in the next sections. 

\begin{figure}
\includegraphics[width=\hsize]{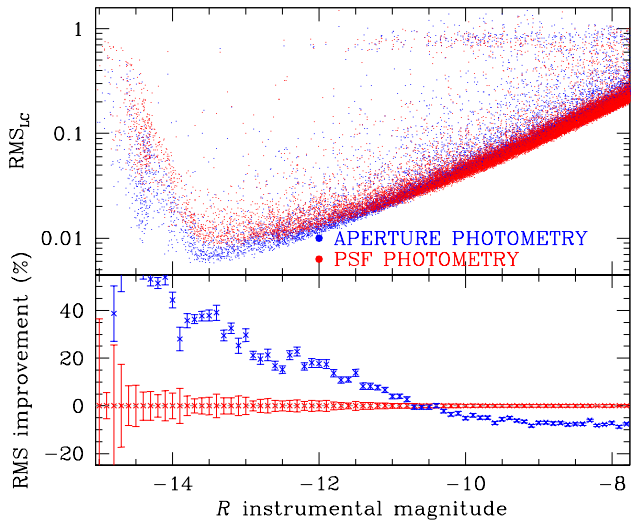}
\caption{
{\it Top panel}: photometric rms as function of $R$ instrumental magnitude for 
PSF photometry (red) and for aperture photometry (blue). 
{\it Bottom panel}: percentage variation with respect to PSF-fitting
photometry. This shows that for bright stars (instrumental $R<-10.5$) 
aperture photometry is better than PSF photometry.  
}
\label{detrend}
\end{figure}

Finally, the RMS of the neighbour-subtracted, LZP-detrended aperture 
vs.~PSF-fitting photometry is compared in Fig.~\ref{detrend}, for long 
exposures in the $R$ filter. It appears that aperture photometry 
performs on average better than PSF-fitting photometry on bright stars 
(with instrumental magnitude $<-10.5$), while PSF-fitting photometry 
gives best results on fainter stars, as expected (see 
Sect. \ref{photwithnonei}). From here on, for each target star in the 
MSL, only the light curve that provides the smallest overall scatter 
goes through the next stages of analysis. 

%%%%%%%%%%%%%%%%%%%%%%%%
%%%%%%%%%%%%%%%%%%%%%%%%
%%%%%%%%%%%%%%%%%%%%%%%%

\section{Variable Finding}
\label{finding}
We employed three different software tools to detect candidate 
variable stars in our dataset of LZP-corrected, neighbour-subtracted 
light curves. These are: the Lomb-Scargle (LS) periodogram 
\citep{1976Ap&SS..39..447L,1982ApJ...263..835S}, the Analysis of 
Variance (AoV) periodogram \citep{1989MNRAS.241..153S} and the 
Box-fitting Least-Squares (BLS) periodogram 
\citep{2002A&A...391..369K}. 
All these tools are implemented within the code \texttt{VARTOOLS} 
v1.202, written by \citet{2008ApJ...675.1254H} and publicly 
available\footnote{http://www.astro.princeton.edu/~jhartman/vartools.html}. 

The LS algorithm is most effective in detecting sinusoidal or 
pseudo-sinusoidal periodic variables. It provides the formal false 
alarm probability (FAP) as a quantitative diagnostic to select stars 
with the highest probability to be genuine variables.  We searched 
for more general types of periodic variables through the AoV 
algorithm, as it is based on variance minimization with phase binning, 
which is a more flexible approach. As for the LS method, we used the AoV 
FAP metric $\Theta$ to select stars that have a high probability to be 
variables. 
The BLS algorithm is particularly effective when searching for 
box-like dips in a otherwise flat or nearly flat light curve, such as 
those caused by detached eclipsing binaries and planetary transits. In 
order to select good candidates, we used the diagnostic parameter 
``signal-to-pink noise'' \citep{2006MNRAS.373..231P}, as defined by 
\citet{2009ApJ...695..336H}.  Before running the periodograms, we 
converted the temporal axis of all light curves from Modified Julian 
Date (MJD) to the Barycentric Julian Date (BJD) reference frame and 
Barycentric Dynamical Time (TDB) standard. This standard provides 
consistency and is much more reliable when comparing observations on 
very long baselines \citep{2010PASP..122..935E}. 
%%%%%%%%%

We searched for variables independently on both the aperture- and the 
PSF-fitting light curves (LZP-detrended, from neighbour-subtracted 
images).  These finding algorithms were also run independently on both 
the $R$ and the $N$ photometric series. 
On the $R$ light curves we employed LS and AoV to search for variables 
with periods between 0.01 and 881 days, with a sampling of 0.0005 
times the Nyquist one in the frequency space. The period interval set 
for the BLS algorithm is between 0.5 and 881 days. On the unfiltered 
light curves ($N$), LS and AOV were set to periods between 0.01--85.5 
days (frequency sampling of 0.0005 times the Nyquist one), while BLS 
between 0.5--85.5 days. 

\begin{figure}
\includegraphics[width=0.95\hsize]{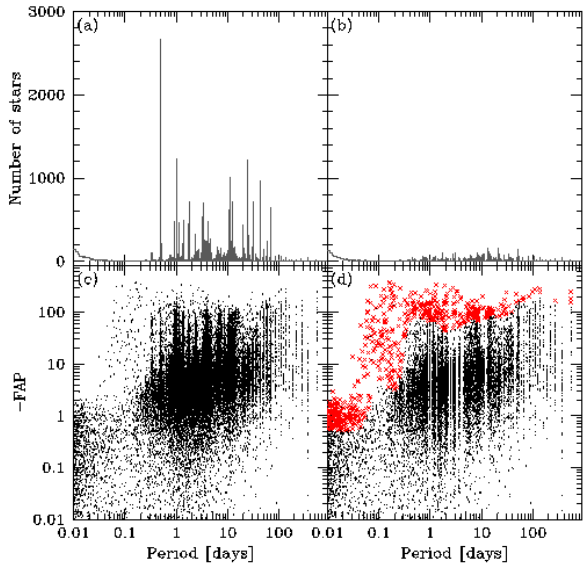}
\caption{
  Example of the procedure followed for the extraction of candidate
  variables from the dataset of light curves in the case of LS
  algorithm. {\it Panel (a)}: distribution of the periods obtained from all
  the light curves; {\it panel (b)}: distribution of the periods after
  spikes suppression; {\it panel (c)}: periods of light curves as a 
  function of FAP for all the stars; {\it panel (d)}: the same of
  panel (c) after spikes suppression. The suspected
  variables are displayed inn red. See text for details.}
\label{finds}
\end{figure}

For all three finding methods (LS, AoV, BLS) we applied the procedure
illustrated in Fig.~\ref{finds} to identify the candidate
variables. We first constructed the histogram of the detected periods
for all the light curves (panel (a) of Fig.~\ref{finds}); spikes in
the histogram at this stage are probably associated to spurious
periods due to systematic errors, such as instrumental and atmospheric
artifacts, which affect a significant fraction of light curves even
after the LZP correction. We removed from our catalogue those stars
close to the spikes 
 as follows: for each period $P_0$ of the histogram of
  Fig.~\ref{finds}a, we considered a region around it of radius equal
  to $50\times \delta P$, where $\delta P$ is the
  bin chosen to build the histogram. We computed the median of the
  counts of the considered bins, and we flagged $P_0$ as spike if the
  associated counts are 5$\sigma$ above the median, where $\sigma$ is
  the 68.27th percentile of the sorted residuals from the median value.
  For the periods associated to spikes, we kept the light curves with low FAP
  (or high AoV or signal-to-pink-noise), i.e. the light curves above the 99.5th
  percentile of the $-$FAP (or the AoV or signal-to-pink-noise -- see
    panel (b) of Fig.~\ref{finds}).
Panels (c) and (d) show the FAP as a function of the 
detected period, respectively before and after the spike 
subtraction. As a last step to select candidate variables, we divided 
the distribution of panel (d) in 25 period bins and selected only 
stars above the 97th percentile of $-$FAP. 

Finally, we combined the lists of candidates obtained with the three 
aforementioned methods, and visually inspected each of them.  We 
identified 519 real variables: $BVRN$ light curves are available for 
442 of them, while for 37 objects only $BVR$ light curves 
are available.  Nineteen stars have only $N$ light curves, and the 
remaining 21 have only short exposures in $R$. 
For the 442 variables in common between $R$ and $N$, we refined their 
periods with the following recipe. For each star we normalized the $R$ 
and the $N$ light curves to zero by subtracting the median 
magnitude. Then we merged the two light curves. In this way we 
obtained a normalized light curve that spans three years. 
We ran the VARTOOLS algorithms LS, AoV, and BLS on this
  normalized light curve using the same parameters described above, to
  find the best period of the associated variable star.
While no modelling can be carried out on such hybrid light curves, the
uncertainty on the resulting period is clearly lowered.

%%%%%%%%%%%%%%%%%%%%%%%%
%%%%%%%%%%%%%%%%%%%%%%%%
%%%%%%%%%%%%%%%%%%%%%%%%

\section{Variables and Light Curves}
\label{variables}

Amongst the 66\,486 stars we found 519 variables, which appear to be 
point-like sources in the $R$ and $N$ stacked images: 246 of them are already 
known variable stars and 273 are new ones. The list of all variables 
is given in Table~\ref{tab:var}: we provide our identification number, 
position, period, magnitudes (when available) in $BVRJ_{\rm 2MASS}H_{\rm 
  2MASS}K_{\rm 2MASS}$, variable type\footnote{The classification of 
  variable stars was done by eye, and sometimes was difficult.} 
and the cross-identification with known variables. 

\begin{figure*}
\includegraphics[bb=36 172 588 712, width=\hsize]{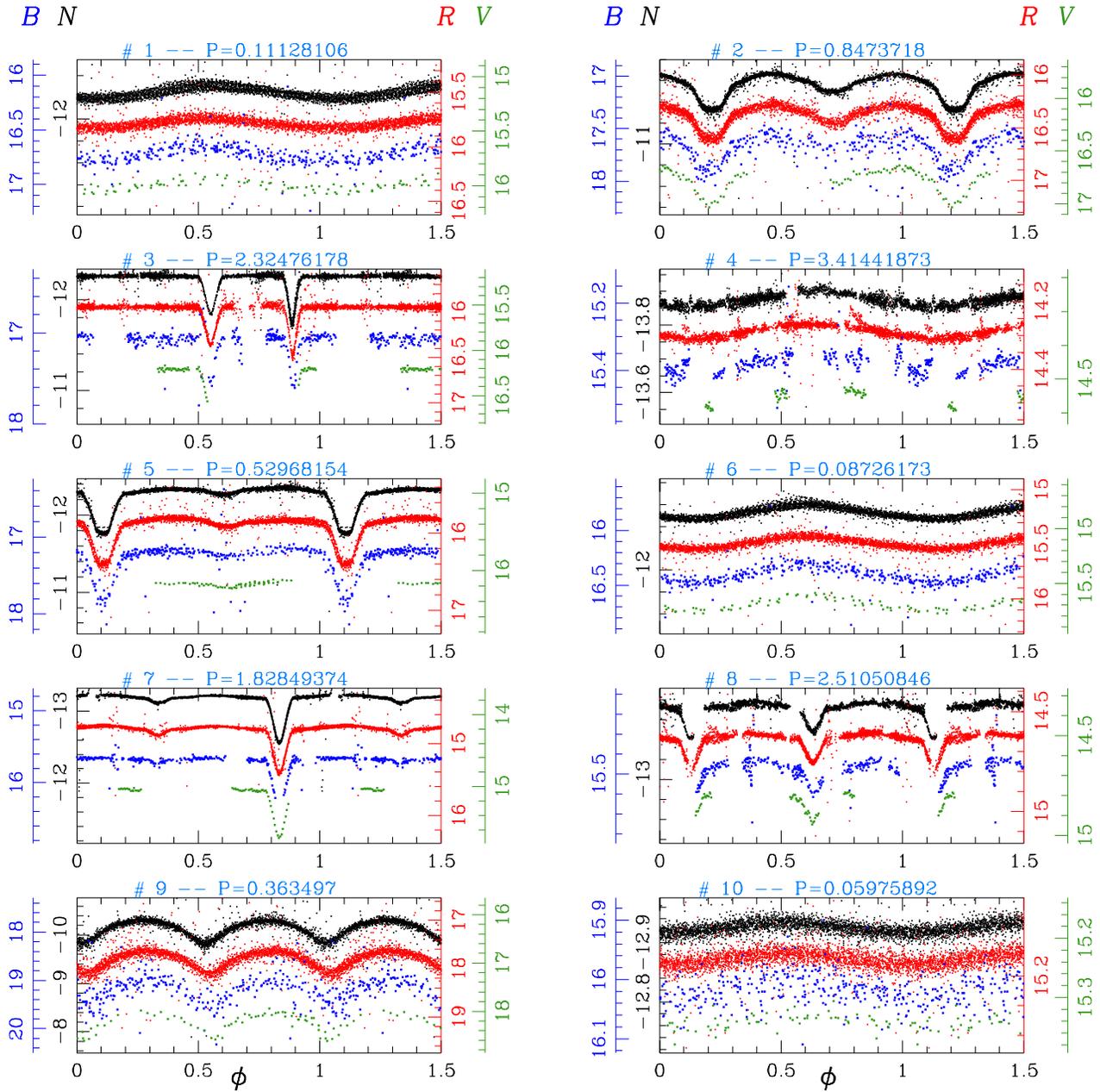}
\caption{
  Light curves of the first 10 variables. For each variable star we
  show (when available) the $N$ (black), $R$ (red), $B$ (blue) and $V$
  (green) light curves from top to bottom of each panel. } 
\label{lcs}
\end{figure*}

Figure~\ref{lcs} shows the folded light curves of all the variables 
listed in Table ~\ref{tab:var} (all figures are available in the 
electronic version of the journal): each panel shows, from 
top to bottom, the light curves in white light $N$ (black), in $R$ 
(red), $B$ (blue) and $V$ (green) filters. For each filter the y-axis 
has the same extension in magnitude range.  The identification 
number and the determined period are reported above each panel.  

\begin{figure*}
\includegraphics[bb=22 160 441 456, width=\hsize]{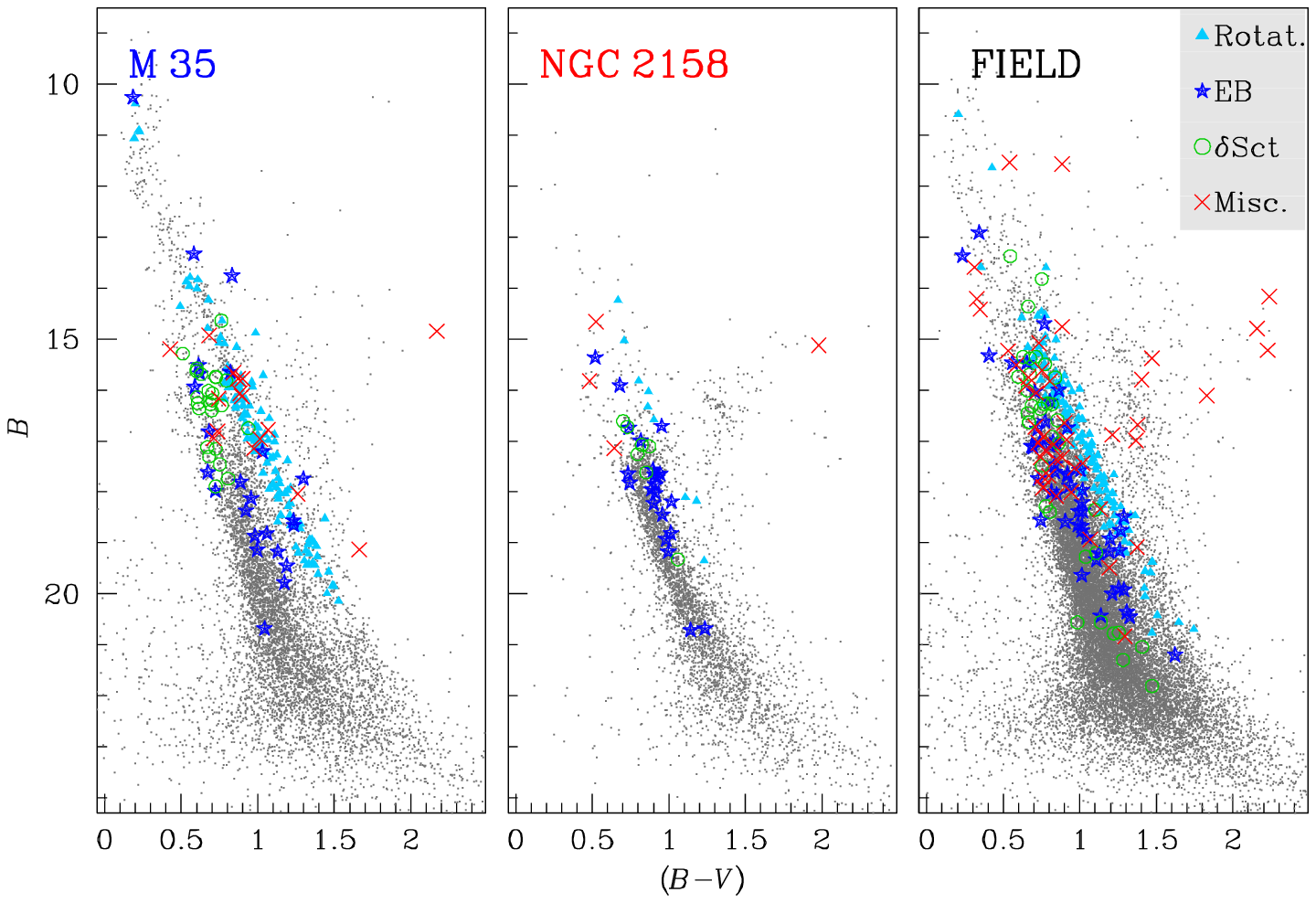}
\caption{
  {\it Left panel}: stars inside the region delimited by the blue 
  circle (M~35) in Fig. \ref{fov}.  {\it Central panel}: stars inside 
  the region delimited by the red circle (NGC~2158) in Fig. \ref{fov}. 
  {\it Right panel}: stars outside the region delimited by the blue 
  and red circles in Fig. \ref{fov}.  
  Different symbols represent different variable types: azure 
  triangles for rotating stars, blue stars for eclipsing binaries, 
  green open circle for $\delta$ Scuti and red crosses for the other  
  variables.  }
\label{cmd}
\end{figure*}

Figure~\ref{fov} shows the stacked image in the $R$ filter.  For 
illustrative purposes we define two regions around the two clusters: 
the blue circle which should contain prevalently stars belonging to 
M\,35 inside an arbitrary radius of about $13\farcm8$, while the red  
circle, with a radius of $6\farcm$0 should include mainly NGC\,2158 
stars. The $B$ vs. $(B-V)$ color-magnitude diagram (CMD) of the stars 
within the blue circle is shown in the left panel of Fig.~\ref{cmd}. 
The central panel shows the CMD of the stars within the red circle and 
in the right panel the CMD of the stars outside both regions. Clearly, 
the CMD of each region shows contamination by stars from the other two 
regions. 
%%%%%%%%%%%%%%%%%%%%%%%%%%%%%%%%%%%%%%%%%%%%%%%%%%%%%%%%%%%%%%%%%%%

\subsection{Membership} 
We used the membership probabilities given by 
\citet{2014A&A...564A..79D} to verify the membership of the identified 
variable stars. We found 248 common stars between our catalogue and 
their M\,35 catalogue; of these stars, 197 have a probability $\geq 
50\%$ to belong to M\,35.  By matching our catalogue of variable 
stars with the NGC\,2158 catalogue of \citet{2014A&A...564A..79D}, we 
found only 9 stars in common, 5 of which have a membership probability 
$\geq 50\%$.

The membership probabilities for the two clusters are tabulated in 
columns 12 and 13 of Table~\ref{tab:var}. 
%%%%%%%%%%%%%%%%%%%%%%

\begin{table*}
\caption{First ten lines of the catalogue of variable stars.}
\tiny
\medskip
\label{tab:var}
\begin{tabular}{l c c c c c c c c c c c c c c}
\hline
N & $\alpha$(J2000) & $\delta$(J2000) & $P$(days) & $B$ & $V$ & $R$ & $N$ & $J_{\rm 2MASS}$  & $H_{\rm 2MASS}$  & $K_{\rm 2MASS}$ & Type & MP$_{\rm M35}^{7}$ & MP$_{\rm NGC2158}^{7}$ & Notes \\
(1) & (2) & (3) & (4) & (5) & (6) & (7) & (8)  & (9)  & (10) & (11) & (12) & (13) & (14) & (15-20) \\
\hline
\hline
  1     & 91.858422   & +24.073243   & 0.11128106     & 16.72   & 15.99   & 15.78 & -11.98  & 14.316     & 14.133    & 13.731 & $\delta$Sct    &-99.999& 62     &                       V37$^3$                                               \\
  2     & 92.561694   & +24.006494   & 0.8473718      & 17.64   & 16.72   & 16.35 & -11.32  & 14.749     & 14.420    & 14.230 & EB             &-99.999& -99.999&                                                                             \\
  3     & 92.488468   & +24.058195   & 2.32476178     & 17.06   & 16.21   & 15.94 & -11.77  & 14.546     & 14.286    & 14.066 & EB             &-99.999& -99.999&                                                                             \\
  4     & 91.858693   & +24.43994    & 3.41441873     & 15.37   & 14.54   & 14.28 & -13.44  & 12.974     & 12.562    & 12.478 & Rot            &-99.999& -99.999&                                                                             \\
  5     & 92.335142   & +24.254324   & 0.52968154     & 17.20   & 16.17   & 15.85 & -11.82  & 14.634     & 14.014    & 13.925 & EB             & 96    & -99.999&          V7 $^2$                                                         \\
  6     & 92.350003   & +24.314434   & 0.08726173     & 16.40   & 15.70   & 15.48 & -12.30  & 14.197     & 13.888    & 13.862 & $\delta$Sct    & 0     & -99.999&                                                   V16$^5$                   \\
  7     & 92.221929   & +24.477042   & 1.82849374     & 15.67   & 15.05   & 14.79 & -12.92  & 13.491     & 13.094    & 12.996 & EB             & 94    & -99.999&                                                   V15$^5$                   \\
  8     & 92.058242   & +24.510734   & 2.51050846     & 15.47   & 14.81   & 14.62 & -13.14  & 13.403     & 13.153    & 13.062 & EB             & 96    & -99.999&                                                                             \\
  9     & 91.919161   & +24.084334   & 0.363497       & 19.16   & 18.17   & 17.77 &  -9.86  & 16.104     & 15.669    & 15.293 & EB             &-99.999& -99.999&                       V05$^3$,        25 $^4$                                \\
  10    & 92.106682   & +24.257778   & 0.05975892     & 16.01   & 15.33   & 15.17 & -12.63  & 14.081     & 13.897    & 13.718 & $\delta$Sct    & 95    & -99.999&                                                                             \\
\hline
\end{tabular}

\justify
\small
{\bf Notes.} \\
$^1$~\citet{2004IBVS.5558....1K}; 
$^2$~\citet{2005ChJAA...5..356H}; 
$^3$~\citet{2006AJ....131.1090M}; 
$^4$~\citet{2009ApJ...695..679M}; 
$^5$~\citet{2010PKAS...25..167.};
$^6$~GCVS \\
$^7$~Membership Probabilities for M~35 and NGC~2158 expressed in \% : 
http://www.astro.iag.usp.br/ocdb/ \citep{2014A&A...564A..79D}
\end{table*}

 Figure~\ref{ccmd} shows two CMDs with the largest colour baseline, 
 the CMD in the Johnson filters, $B$ 
 versus $(B-R)$ (left panel), and the CMD obtained combining the white 
 light $N$ (which peaks at $\sim$500\,nm) and the 2MASS IR-filters, 
 $J_{\rm 2MASS}$ versus $(N-K_{\rm 2MASS})$ (right panel). 
In both CMDs we plot the variable stars we have identified, with
different symbols and colours according to their type.

\subsection{Eclipsing binaries}

\begin{figure}
 \centering
\includegraphics[width=0.5\textwidth]{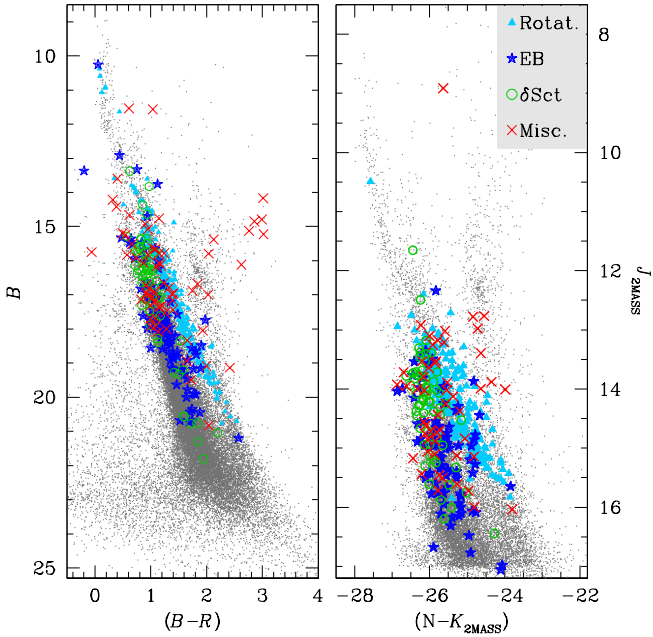}
\caption{ $B$ vs. ($B$$-$$R$) ({\it left panel}) and $J_{\rm 2MASS}$ 
  vs. ($N$-$K_{\rm 2MASS}$) ({\it right panel}) CMDs. The variable 
  stars are displayed as follows: azure triangles are rotational 
  stars, blue stars are eclipsing binaries, green circles are 
  $\delta$~Sct and red crosses are other variables such as long 
  period, nonperiodic, etc. }
\label{ccmd}
\end{figure}

We extracted 97 light curves of eclipsing binaries (EBs): 41 of these 
were previously known, the others are new.  Many of them are detached 
systems that, if cluster members, offer the potential for obtaining 
very precise cluster ages and distances and for testing stellar evolution  
models, e.g., along the lines of \citet{2012A&A...543A.106B}.  For 9 of 
these new EBs we have membership probabilities from 
\citet{2014A&A...564A..79D}: we found that 8 of them have a high 
probability ($\geq 87\%$) to be members of M\,35. They are V8, V149, 
V231, V271, V422, V513, V514 and V517. For NGC\,2158 we have no 
membership information, but looking at Fig.~\ref{ccmd} there is a good 
chance that a substantial fraction of the eclipsing systems in the 
field of NGC\,2158 are members, since they are located along the 
cluster sequence in the CMD. 

The period distribution of the identified EBs is likely to be 
significantly biased towards short periods and the periods for the 
longer period systems are uncertain as in common for ground based 
surveys (\citealt{1996MNRAS.282..705R}). However, complementary 
observations from {\it K2} should alleviate these issues once 
analyzed.

\subsection{Rotating Stars}
We classify 284 of our variables as rotating stars, of which 122 are 
new discoveries. 
Their light curves show sinusoidal light variations, mainly due to the 
starspots that follow the surface rotation. We found that, for many 
of these stars, the shape of the light curves changes during the three years, 
but their period remained fairly unchanged (see V18 for an 
example). Other stars show a sinusoidal shape light curve in one  
observational run, but their light curve is flat in the other  
runs (for example V237). There are stars that changed their median 
magnitude from one year to another, such as V274 or V302. 

\begin{figure}
\centering
\includegraphics[width=0.5\textwidth]{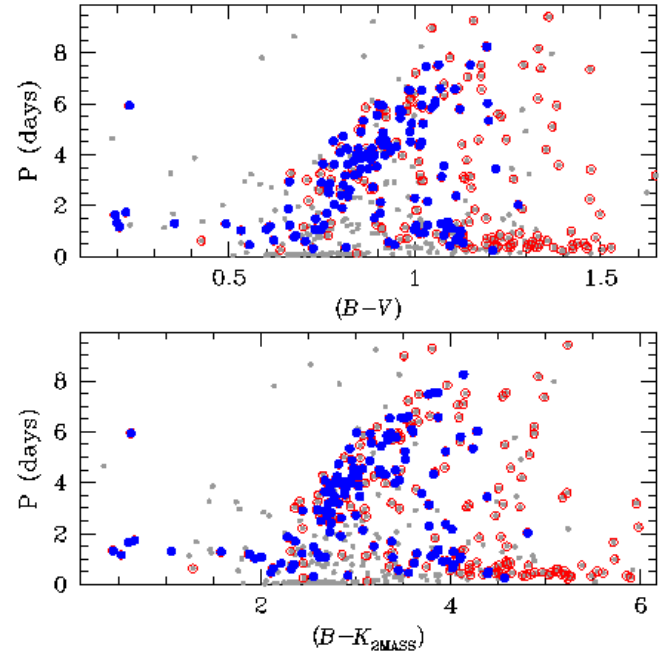}
\caption{
  Period vs ($B$$-$$V$) ({\it top panel}) and 
  ($B$-$K_{\rm 2MASS}$) ({\it bottom panel}) for all the  
  variables (grey dots). Red circles mark the rotating stars, that define 
  the colour-period relation for the stars of M\,35. 
  The rotating stars with a membership probability $\geq 50\%$ are displayed in blue. }
\label{colper}
\end{figure}

Figure~\ref{colper} shows the relation between the period of the 
variable stars and their $B-V$ (top panel) and $B-K_{\rm 2MASS}$ 
(bottom panel) colours: the variable stars are in grey, red circles 
represent the variable stars identified as rotating stars, while 
blue dots are rotating stars with a probability $\geq 50\%$ to be 
M\,35 members. 

\citet{2009ApJ...695..679M} carried out a detailed analysis of the  
period-$(B-V)$ colour index relation using 310 stars members of M\,35, of which 
153 are in common with our catalogue.

\subsection{$\delta$ Scuti and other variable stars}

We identified 67 $\delta$ Scuti stars, 45 of which 
were previously unknown. In the CMDs, they are mainly located above 
the main-sequence turn-off of NGC\,2158. 

Finally, we found 69 variable stars that show long period variations 
or nonperiodic signals, 21 of these already known. We also found a 
$\delta$~Cep (V442, already known) and a RRLy (V492).

% 

%%%%%%%%%%%%%%%%%%%%%%%%
%%%%%%%%%%%%%%%%%%%%%%%%
%%%%%%%%%%%%%%%%%%%%%%%%

\section{The Electronic Material}
\label{emat}

The catalogue of all variable stars is available electronically (or 
upon request to the authors). The catalogue contains the following 
information: Col.\,(1) displays is the identification number of the variable 
in this work; Cols.\,(2) and (3) are the J2000 equatorial coordinates 
in decimal degrees; Col.\,(4) contains the period in days; Cols.\,(5) 
to (11) are the calibrated $BVR$, instrumental $N$ and calibrated 
$J_{\rm 2MASS}H_{\rm 2MASS}K_{\rm 2MASS}$ magnitudes (when magnitudes 
are not available in a particular filter, they are denoted as 0.0); 
Col\,(12) provides the variable type; Col\,(13) and (14) give the 
membership probabilities for M\,35 and NGC\,2158 stars from 
\citet{2014A&A...564A..79D} (when the membership probability is not 
available, it is denoted as -99.999); Cols.\,(15)--(20) give the 
identification ID in other published catalogues, specifically: 
\citealt{2004IBVS.5558....1K}, \citealt{2005ChJAA...5..356H}, 
\citealt{2006AJ....131.1090M}, \citealt{2009ApJ...695..679M}, 
\citealt{2010PKAS...25..167.} and GCVS (in Table~\ref{tab:var} we 
combined cols. (15)--(20) to best fit the table in the manuscript). 

For each star in the catalogue we release an astrometrized 
$43\farcs1 \times 43\farcs1$ finding chart for each available filter. 

A catalogue of all sources detected in $N$ and $R$ filters is also 
electronically available. In this catalogue, Cols.\,(1) and (2) are the 
J2000 equatorial coordinates in decimal degrees; Cols.\,(3)--(9) are 
the calibrated $BVRJ_{\rm 2MASS}H_{\rm 2MASS}K_{\rm 2MASS}$ magnitudes 
and the instrumental $N$ magnitudes. As mentioned in Sect.~\ref{stack}, 
the astrometrized stacks in $BVRN$ filters are also electronically available.

\section{Summary}
\label{summary}
We present the first results of a long term photometric survey of OC 
stars conducted with the 67/92 cm Schmidt telescope at Cima Ekar, 
Asiago. In this paper, we focus on a field which 
includes the two open clusters M35 and NGC 2158.  A total of 6996 
58$\times$38 arcmin$^2$ images in $B$, $V$, $R$, and white light (no 
filter) were collected over 2.4 years. We test four different 
approaches to the stellar photometry: aperture and PSF 
photometry on the original and neighbour-subtracted images. Aperture 
photometry on neighbour-subtracted images proves to be the most 
appropriate method to obtain light curves with the lowest photometric RMS. We 
also test different approaches to correct our photometry for 
systematic errors.  Our final database includes 66\,486 stars.  We run 
different algorithms for the identification of variable stars: i) the 
Lomb-Scargle periodogram, ii) the Analysis of Variance periodogram, 
and iii) the Box-finding Least Square periodogram. We identify 519 
variables: 97 eclipsing binaries (56 are new identifications), 284 
rotational variables (122 new), 67 $\delta$~Scuti variables (45 are 
new), 69 long period or non periodic variables (50 of them are new), 1 
RRLy and 1 $\delta$~Cep.  Finally, we cross-correlate our variable 
star catalogue with previously published catalogues to obtain 
membership probabilities, and for 
cross-identification with already know variables.  The catalogue with 
coordinates, $B$, $V$, $R$, 2MASS magnitudes, membership and 
cross-identification with known variables is made electronically 
available. The electronic material includes the $B$, $V$, $R$ and 
white light stacked images. For each variable, a 43.1$\times$43.1 
arcsec$^2$ finding chart is also made available. 
M\,35 and NGC\,2158 are within the field of {\it Campaign-0} of the
{\it Kepler\,2 Mission}. Our survey will therefore complement ---and extend
in time--- the light curves of variable (and not variable) stars
covered by K2.
%%%%%%%%%%%%%%%%%%%%%%%%
%%%%%%%%%%%%%%%%%%%%%%%%
%%%%%%%%%%%%%%%%%%%%%%%%

%%%%%%%%%%%%%%%%%%%%%%%%

\section*{Acknowledgements}

D.N., L.R.B., V.N., M.L., A.C., G.P., L.B., V.G., and L.M. acknowledge
PRIN-INAF 2012 partial funding under the project entitled: ``The M4
Core Project with Hubble Space Telescope''. D.N. is supported by a
grant ``Borsa di studio per l'estero, bando 2013'' awarded by
``Fondazione Ing.\,Aldo Gini'' in Padua (Italy).
Some tasks of our data analysis have been carried out with the
  VARTOOLS \citep{2008ApJ...675.1254H} and ASTROMETRY.NET codes \citep{2010AJ....139.1782L}. This research made use of the International Variable Star
  Index (VSX) data base, operated at AAVSO, Cambridge, MA, USA.
%%%%%%%%%%%%%%%%%%%%%%%%%%%%%%%%%%%%%%%%%%%%%%%%%%%%%%

\bibliographystyle{mn2e} 
\bibliography{biblio}
\bsp

\label{lastpage}

\end{document}